\documentstyle[aps,prb,epsfig,twocolumn]{revtex}
\begin{document}

\draft

\twocolumn[
\hsize\textwidth\columnwidth\hsize\csname @twocolumnfalse\endcsname

\title{Impurity effects on s+g-wave superconductivity in
borocarbides Y(Lu)Ni$_2$B$_2$C}
\author{Qingshan Yuan,$^{1,2}$ Hong-Yi Chen,$^1$ H. Won,$^3$ S. Lee,$^3$ 
K. Maki,$^4$ P. Thalmeier,$^5$ and C. S. Ting$^1$}
\address{$^1$ Texas Center for Superconductivity and Department of Physics,
University of Houston, Houston, TX 77204\\
$^2$Pohl Institute of Solid State Physics, Tongji University,
Shanghai 200092, P.R.China\\
$^3$ Department of Physics, Hallym University, Chunchon 200-702, South Korea\\
$^4$ Department of Physics and Astronomy, University of Southern California,
Los Angeles, CA 90089-0484\\
$^5$ Max-Planck-Institut f\"ur Chemische Physik fester Stoffe,
N\"othnitzer Str. 40, 01187 Dresden, Germany\\
}

\maketitle

\begin{abstract}
Recently a hybrid s+g-wave pairing is proposed to describe the experimental 
observation for a nodal structure of the superconducting gap in borocarbide 
YNi$_2$B$_2$C and possibly LuNi$_2$B$_2$C. In this paper the impurity
effects on the s+g-wave superconductivity are studied in both Born and unitarity limit. 
The quasiparticle density of states and thermodynamics are calculated. 
It is found that the nodal excitations in the clean system are immediately
prohibited by impurity scattering and a finite energy gap increases quickly with 
the impurity scattering rate. This leads to an activated behavior in the temperature 
dependence of the specific heat. Qualitative agreement with the experimental results 
is shown. Comparison with d-wave and some anisotropic s-wave studied previously
is also made.

\end{abstract}

\pacs{PACS numbers: 74.20.Rp, 74.62.Dh, 74.70.Dd}
]

\section{Introduction}

The superconductivity in rare earth nickel borocarbides RNi$_2$B$_2$C
(R=Y, Lu, Tm, Er, Ho, and Dy) is of great interest in recent years.
\cite{Canfield,Muller} Among them, the two nonmagnetic borocarbides 
YNi$_2$B$_2$C and LuNi$_2$B$_2$C have relatively high superconducting (SC)
transition temperatures, which are 15.4K and 16.5K, respectively.
Although they were initially understood by an isotropic s-wave 
pairing,\cite{Carter} recent various experimental results, including 
specific heat,\cite{Nohara97,Nohara99,Izawa01,Park}
thermal conductivity,\cite{Izawa02,Boaknin} Raman scattering,\cite{Yang,Lee} 
NMR relaxation rate,\cite{Zheng} photoemission spectroscopy,\cite{Yokoya}
scanning tunneling microscopy and spectroscopy,\cite{Samper}
have shown that they may be another class of nodal superconductors.
For example, both the $\sqrt{H}$ dependence of the specific heat 
and $H$ linear dependence of the thermal conductivity in the vortex state 
indicate the existence of nodal excitations.\cite{Nohara97,Nohara99,Izawa01} 
In particular, compelling evidences are presented by Izawa {\it et al.}\cite{Izawa02}
based on the angular dependent thermal conductivity measurements in a
magnetic field that the gap function of YNi$_2$B$_2$C is highly anisotropic with
anisotropy ratio less than $10^{-2}$, i.e., it has essentially nodes and
moreover, the nodes are point-like which are 
located along [1,0,0] and [0,1,0] directions.
Due to Doppler shift in the presence of supercurrent flow,\cite{Volovik} 
the zero energy quasi-particle density of states (DOS) becomes finite
and is sensitive to the relative orientation 
between the magnetic field and nodal direction.
Thus the nodes can be unambiguously extracted from 
the angular modulation of the thermal conductivity.
Based on the same physical mechanism the angular dependent specific heat 
in YNi$_2$B$_2$C was measured by Park {\it et al.}\cite{Park} 
and the conclusion for the existence of the point nodes can be also drawn.
In addition, the thermal conductivity in LuNi$_2$B$_2$C
was measured by Boaknin {\it et al.} as a function of temperature and 
field strength.\cite{Boaknin} They claimed that
the gap minimum is at least $10$ times smaller than the gap maximum
and possibly goes to zero at nodes.
Thus YNi$_2$B$_2$C (and presumably LuNi$_2$B$_2$C) becomes the first 
superconductor in which the gap function with point nodes 
is clearly identified.

Theoretically, Maki {\it et al.} have proposed that the so called s+g-wave
spin singlet gap function, i.e.,
\cite{Maki,Thalmeier,Izawa02}
\begin{equation}
\Delta_{\bf k}=\Delta (1 - \sin ^4\theta \cos 4\phi)/2
=\Delta f({\bf k})\ ,
\label{gapfun0}
\end{equation}
can describe the experimental observation in Y(Lu)Ni$_2$B$_2$C. 
Here $\theta, \phi$ are the polar and azimuthal angles of ${\bf k}$,
respectively. The second `g-wave' component is given by a
fourth degree A$_{1g}$ basis function: $k_x^4+k_y^4-6k_x^2 k_y^2$
in tetragonal crystal symmetry.
In the gap function the amplitudes of s- and g-components are 
assumed to be equal. Thus 4 (and only 4) point nodes at
$\theta=\pi/2$ and $\phi=0, \pi/2, \pi, 3\pi/2$ are realized, 
which is exactly what has been observed experimentally.\cite{Izawa02}

So far, a microscopic theory for the s+g-wave pairing
is not available which might be complicated due to the
complex Fermi surface of borocarbides\cite{Dugdale} and the
possibility of strongly anisotropic Coulomb interactions.
A phenomenological theory has been constructed by two of us,\cite{Yuan}
which shows that a stable coexistence of s- and g-wave with
the fine tuning, i.e., the equal amplitudes of them are possible.
On the other hand, the correctness of the s+g model could be checked
by comparing the theoretical predictions with the available experimental data. 
Up to now, quite a few physical quantities based on the s+g model
such as thermal conductivity,\cite{Izawa02} Raman spectra,\cite{Jang}
sound attenuation\cite{Won} etc. have been calculated and 
good agreement with the experimental results is obtained.
This in turn stimulates us to continue studying more properties.

In this work the impurity effects on the s+g model will be investigated.
Experimentally there have been extensive studies on Pt-substituted
borocarbide Y(Ni$_{1-x}$Pt$_x$)$_2$B$_2$C.
The ultrahigh-resolution photoemission spectroscopy
was performed by Yokoya {\it et al.}\cite{Yokoya}
They found a small but significant difference for the spectra
in the SC states between $x=0$ and $x=0.2$ compound, 
which is contributed to the change in anisotropy of the SC gap: 
a strong anisotropic gap in $x=0$ and an almost isotropic one in $x=0.2$. 
That is, the gap anisotropy is wiped out by introducing impurities. 
Consistent result can be obtained from the electronic specific heat $C$ measurements
by Nohara {\it et al.}\cite{Nohara99} 
The temperature-dependence of $C$ (under zero field) shows a power law behavior 
for $x=0$ but a thermally activated behavior for $x=0.2$. Very recently, 
the angular dependent thermal conductivity for Y(Ni$_{1-x}$Pt$_x$)$_2$B$_2$C 
with $x=0.05$ in the SC state was measured.\cite{Kamata} 
There is no angular modulation left. This shows that the gap anisotropy is 
nearly completely suppressed by only $5\%$ Pt substitution.
It is the purpose of this work to see how soon a full energy gap will be
opened by the impurities for the s+g-model through the calculation of the DOS.
Furthermore, the thermodynamics like specific heat will 
be calculated. The crossover from a power law behavior in a pure system
to an activated behavior in the presence of impurities will be exhibited.

\section{Formalism}
The effects of nonmagnetic impurities are treated by the self-consistent
T-matrix approximation.\cite{Hirschfeld} 
For benefit of readers, we briefly summarize the formulation in the following.
To begin with, the Matsubara Green's function in Nambu representation
is written as (quantities with a hat represent matrices)
\begin{eqnarray}
\hat{g}({\bf k},i\omega_n) & = & (i\omega_n\hat{\sigma}_0-\xi_{\bf k}
\hat{\sigma}_3-\Delta_{\bf k}\hat{\sigma}_1-\hat{\Sigma})^{-1}\ ,
\end{eqnarray}
where $\omega_n=(2n+1)\pi T$ ($T$: temperature) is the Matsubara frequency, 
$\xi_{\bf k}$ the electron energy measured from the chemical potential,
$\Delta_{\bf k}$
the gap function which has been assumed to be real.
$\hat{\sigma}_0$ and $\hat{\sigma}_{1,3}$ are unit 
and Pauli matrices, respectively. The self-energy
$\hat{\Sigma}(i\omega_n) = \Gamma_u \hat{T}(i\omega_n)$
with $\Gamma_u=N_i/(\pi N_0)$.
Here $N_i$ is the total number of impurities and $N_0$ is the zero energy 
DOS per spin for the spectrum $\xi_{\bf k}$. The T-matrix $\hat{T}$ satisfies
$\hat{T}(i\omega_n)=\hat{K} + \hat{K}\hat{G}(i\omega_n)\hat{T}(i\omega_n)$,
where $\hat{G}(i\omega_n)=(1/\pi N_0)\sum_{\bf k} \hat{g}({\bf k},i\omega_n)$.
$\hat{K}$ is the T-matrix in the normal state:
$\hat{K}=-\hat{\sigma}_3/c$ with $c=\cot \delta_0$. And $\delta_0$ is the s-wave
phase shift controlled by the strength of a single impurity potential.\cite{Sigrist}

The matrix $\hat{A}$ ($A=\Sigma,T,G$) is decomposed
into $\hat{A}=A_0 \hat{\sigma}_0 + A_1 \hat{\sigma}_1 
+ A_3 \hat{\sigma}_3$. Then one obtains $G_3=0$ if assuming particle-hole
symmetry.\cite{Hirschfeld} Further $T_3$ and correspondingly $\Sigma_3$
vanish if we only concern Born ($c\gg 1$) and unitarity ($c=0$) limit.
The left $\Sigma_{0,1}$ are obtained as follows
\begin{equation}
\Sigma_0 =  \Gamma_u {G_0 \over c^2-G_0^2+G_1^2} \ ,\ \ \
\Sigma_1 =  \Gamma_u {-G_1 \over c^2-G_0^2+G_1^2} \ , \label{Sigma}
\end{equation}
where
\begin{equation}
G_0 = \left\langle {-i\tilde{\omega}_n\over  
\sqrt{\tilde{\omega}_n^2+\tilde{\Delta}_{\bf k}^2}}\right\rangle\ ,\ \ \ 
G_1 = \left\langle {-\tilde{\Delta}_{\bf k}\over
\sqrt{\tilde{\omega}_n^2+\tilde{\Delta}_{\bf k}^2}}\right\rangle\ .
\label{G}
\end{equation}
Above $\tilde{\omega}_n=\omega_n+i\Sigma_0,\ 
\tilde{\Delta}_{\bf k}=\Delta_{\bf k}+\Sigma_1$,
and $\langle\cdots\rangle$ means angular average $\int d\Omega/(4\pi)$.
It deserves to point out that the off-diagonal self-energy $\Sigma_1$ is zero 
for d-wave but finite for the current s+g-wave. 
This difference will lead to different behavior of their respective DOS functions.

The gap function is connected to the matrix element $g_{12}$ by
$\Delta_{\bf k} =  T\sum_{i\omega_n}\sum_{\bf k'} V_{\bf kk'}
g_{12}({\bf k'},i\omega_n)$
with $V_{\bf kk'}$: the pair potential. It is often assumed
$V_{\bf k k'}=-Vf({\bf k})f({\bf k'})$, which is valid 
for $|\omega_n| < \omega_D$ (a cut-off).
Then the following equation for the gap amplitude $\Delta$ is obtained:
\begin{eqnarray}
\Delta & = & 2\pi T \bar{V} \sum_{0<\omega_n < \omega_D}
\left\langle {f({\bf k})(\Delta_{\bf k}+\Sigma_1)\over
\sqrt{(\omega_n+i\Sigma_0)^2+(\Delta_{\bf k}+\Sigma_1)^2}}\right\rangle\ ,
\label{Delta}
\end{eqnarray}
where $\bar{V}=VN_0$ is a dimensionless interaction constant.

By solving Eqs.~(\ref{Sigma})-(\ref{Delta}) self-consistently,
we can obtain the renormalized $\Delta$ with impurity concentrations at
all temperatures and further calculate the observable quantities like
specific heat. In the following we adopt the impurity scattering rate parameter
in the normal state $\Gamma=\Gamma_u/(1+c^2)$, which reduces to
$\Gamma_u/c^2$ in Born limit ($c\gg 1$) and $\Gamma_u$ in unitarity limit ($c=0$).

\section{density of states}

\begin{figure}[ht]
\begin{center}
\epsfig{file=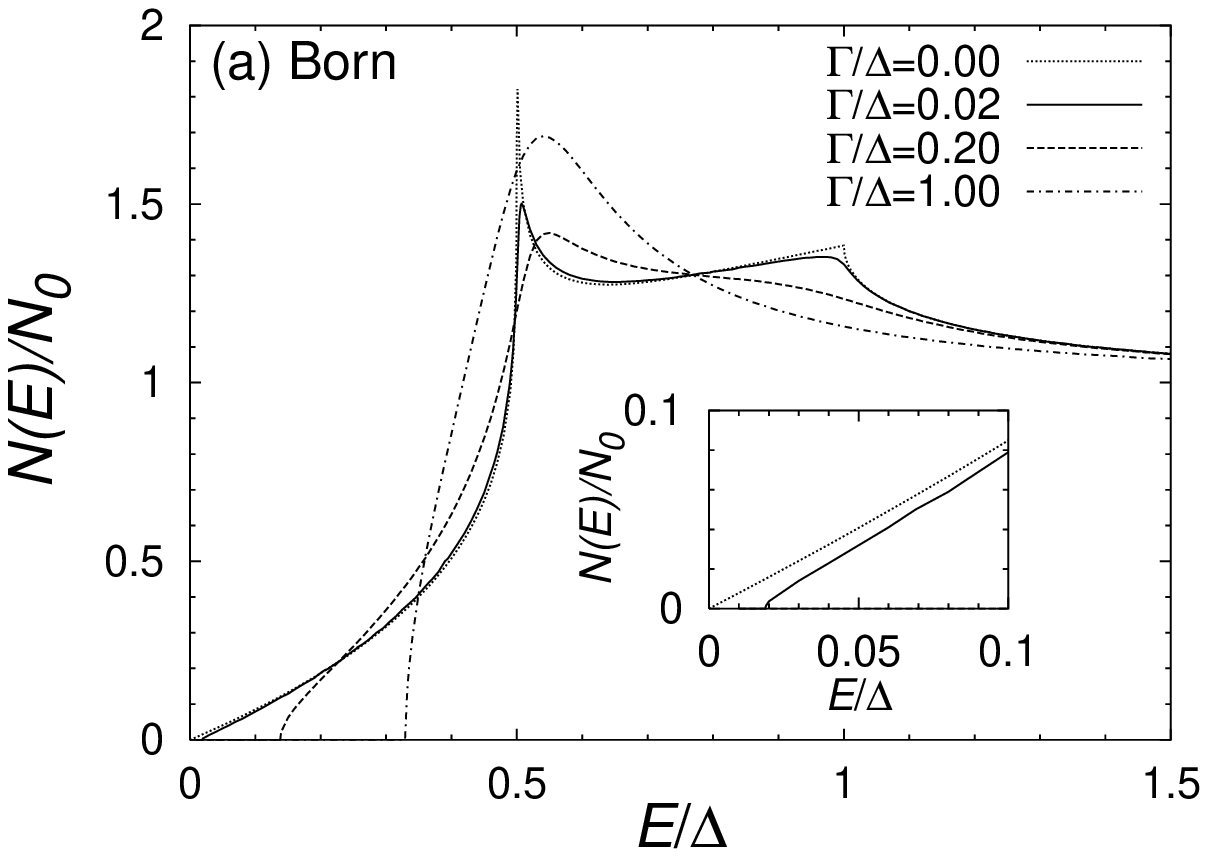,width=8cm,height=6cm}
\end{center}
\begin{center}
\epsfig{file=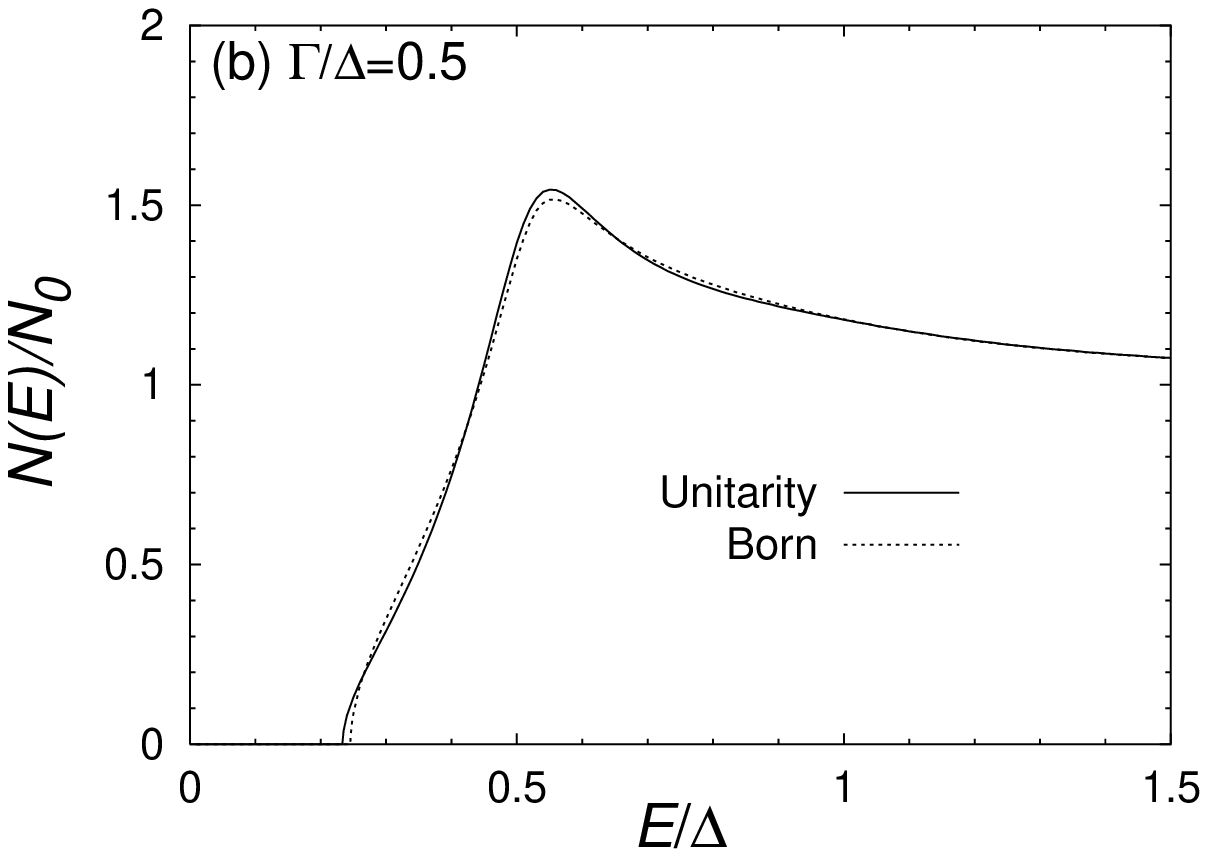,width=8cm,height=6cm}
\end{center}
\medskip
\caption{The quasi-particle DOS for s+g-wave. (a) The results for various $\Gamma$ 
in Born limit. The low energy region is enlarged in the inset. 
(b) Comparison between unitarity and Born limit for a fixed $\Gamma/\Delta=0.5$.
Only subtle differences exist in both limits.}
\label{Fig:DOS}
\end{figure}

We first consider the DOS, which is given by (per spin)
\begin{eqnarray}
N(E)/N_0=-{\rm Im} G_0(i\omega_n)|_{i\omega_n\rightarrow E+i0^+}\ .
\end{eqnarray}
It is noticed that if one scales the input parameters $\Gamma$ and 
$E$ by $\Delta$, the DOS can be obtained by only solving
Eqs.~(\ref{Sigma}) and (\ref{G}) without explicit calculation of $\Delta$ 
for the moment. The results for a few ratios $\Gamma/\Delta$ are shown by 
Fig.~\ref{Fig:DOS} in both Born and unitarity limit. 
(Note that the scaled $\Gamma/\Delta$ is still a monotonous function of $\Gamma$.)
From the figure a few features are clearly seen:
i. There is no essential difference between the results in both limits, as seen by
an example in Fig.~\ref{Fig:DOS}b.
Numerically we have $-G_0^2+G_1^2\simeq 1$ in unitarity limit. Thus
Eq.~(\ref{Sigma}) can be rewritten into a unified form: 
$\Sigma_0=\Gamma G_0,\ \Sigma_1=-\Gamma G_1$ in both limits, leading to
nearly the same DOS.
ii. An energy gap opens up immediately when $\Gamma \neq 0$ and grows quickly 
with $\Gamma$. As seen from the inset of Fig.~\ref{Fig:DOS}a, 
a relatively big energy gap $\omega_g\sim \Gamma$ is already present 
for small $\Gamma/\Delta=0.02$. Actually $\omega_g$ is well approximated by 
$\omega_g=\Gamma/(1+2\Gamma/\Delta)$.
For intermediate $\Gamma/\Delta=1$, the gap anisotropy is nearly suppressed
and the DOS approaches the familiar shape of an isotropic s-wave gap.
All the above results are quite different from those for d-wave.
\cite{Hotta,BH,FN,Sun1,Sun2,PM,HA} 
For example, a finite DOS is induced by impurities around zero energy for d-wave.
Also, a resonance behavior at low energy exists in unitarity limit 
for d-wave, which does not appear in the s+g-state.
We point out that the main result seen from the DOS based on the s+g model,
i.e., the gap anisotropy is suppressed by impurity scattering, is consistent 
with the conclusion drawn from the photoemission data.\cite{Yokoya}

At this stage, we would supplement a remark. 
Previously, in order to clarify the role of sign change in a gap function, 
some {\it hypothetical} anisotropic s-wave gap functions 
e.g. $|\cos 2\phi|$,\cite{BH} 
which vanishes at the same nodes as d-wave but never change sign,
have been studied in the presence of impurities. The obtained DOS 
for $\Gamma\neq 0$ has qualitatively similar properties to that shown here.
This is understandable in the sense that the current s+g-wave (without sign change) 
is essentially an anisotropic s-wave. On the other hand, however,
we would point out an important difference between
the current s+g-wave gap and the one $\sim |\cos 2\phi|$:
The former contributes only a few point nodes, while the latter
gives line nodes, on a three-dimensional Fermi surface.
This will lead to notable differences on some observable quantities, 
e.g., the angular dependent thermal conductivity,
which has actually been used to identify whether the SC gap in YNi$_2$B$_2$C 
has really point or line nodes.\cite{Izawa02}
As far as the DOS is concerned, a quantitative difference
is that the opened gap $\omega_g$ increases with $\Gamma$ for
s+g-wave much faster than that for $|\cos 2\phi|$ since
point nodes are easily removed by impurity scattering.
This is consistent with the experimental observation that YNi$_2$B$_2$C 
in the SC state is very sensitive to impurities as mentioned above. 

\section{transition temperature, SC order parameter and thermodynamics}
In this section we focus on the thorough solution of 
Eqs.~(\ref{Sigma})-(\ref{Delta}). 
Considering the limit $\Delta\rightarrow 0$, we first 
extract the transition temperature $T_c$ analytically. In both Born and 
unitarity limit, we obtain the following equation which is different from the standard
Abrikosov-Gor'kov (AG) one\cite{AG}
\begin{eqnarray}
-\ln {T_c\over T_{c0}} & = & {\langle f^2({\bf k})\rangle-
\langle f({\bf k})\rangle^2 \over \langle f^2({\bf k})\rangle}
\left[\psi ({1\over 2}+{\Gamma\over 2\pi T_c})-\psi ({1\over 2})\right]\ ,\label{Tc}
\end{eqnarray}
where $T_{c0}$ is the transition temperature at $\Gamma=0$, and 
$\psi (z)$ is the digamma function. The prefactor
$p={\langle f^2({\bf k})\rangle-
\langle f({\bf k})\rangle^2 \over \langle f^2({\bf k})\rangle}
\simeq 0.169$ for the current s+g-wave. Note that for the usual nodal 
superconductors with only a single angular momentum pairing like d-wave or pure
g-wave, the prefactor is unity, i.e., Eq.~(\ref{Tc}) becomes the
standard AG equation.\cite{FN,Sun1}
Therefore, for d-wave there exists a critical $\Gamma$:
$\Gamma_c=\pi T_{c0}/(2e^{\gamma})\simeq 0.882 T_{c0}$ at which 
$T_c$ vanishes, whereas for s+g-wave $\Gamma_c=\infty$. In the limit
$\Gamma\rightarrow 0$, Eq.~(\ref{Tc}) gives
\begin{eqnarray}
T_c & = & T_{c0} - {\pi \over 4} p \Gamma \ .\label{Tc:G0}
\end{eqnarray}
\begin{figure}[ht]
\begin{center}
\epsfig{file=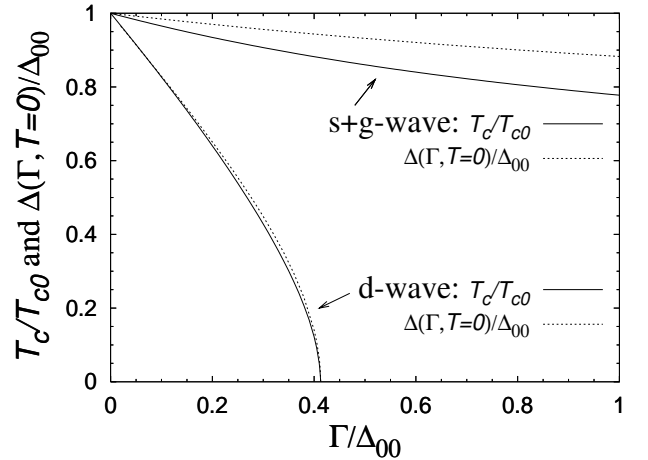,width=8cm,height=6cm}
\end{center}
\medskip
\caption{$T_c/T_{c0}$ (solid lines) and $\Delta(\Gamma,T=0)/\Delta_{00}$ 
(dotted lines) vs. $\Gamma/\Delta_{00}$ for s+g- and d-wave. Note that for each
quantity the results in two limits are the same for s+g-wave. For d-wave
the quantity $\Delta(\Gamma,T=0)/\Delta_{00}$ is shown in unitarity limit.\cite{Sun1}}
\label{Fig:Tc}
\end{figure}
The dependence of $T_c$ on $\Gamma$ is shown in Fig.~\ref{Fig:Tc}, 
where $\Gamma$ is scaled by $\Delta_{00}=\Delta (\Gamma=0,T=0)$. 
The result for d-wave is also shown for comparision.
From Eq.~(\ref{Tc:G0}) we have the linear relation at $\Gamma\rightarrow 0$:
$T_c/T_{c0}=1-q\Gamma/\Delta_{00}$ with $q=0.366$ for s+g-wave and
$1.68$ for d-wave. Here we have used the BCS ratios $\Delta_{00}/T_{c0}\simeq 2.76$
and $2.14$ for s+g- and d-wave, respectively.

We try to evaluate $\Gamma$ from the experimental data on 
$T_c$. For Y(Ni$_{1-x}$Pt$_x$)$_2$B$_2$C, $T_c=12.1$K at $x=0.2$,\cite{Yokoya}
i.e., $T_c/T_{c0}\simeq 0.786$. Then we have $\Gamma/\Delta_{00}\simeq 0.94$.
For $x=0.05$, $T_c=13.1$K,\cite{Kamata} corresponding to 
$\Gamma/\Delta_{00}\simeq 0.55$.

\begin{figure}[ht]
\begin{center}
\epsfig{file=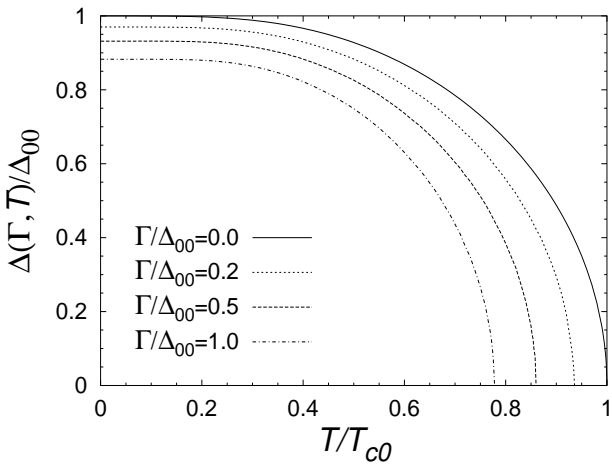,width=8cm,height=6cm}
\end{center}
\medskip
\caption{$\Delta(\Gamma,T)/\Delta_{00}$ as a function of $T/T_{c0}$ at different
$\Gamma/\Delta_{00}$ for s+g-wave.}
\label{Fig:DelT}
\end{figure}

The SC order parameter $\Delta(\Gamma,T)$ is solved numerically.
In what follows we give the results only in Born limit. Those
in unitarity limit are practically the same. 
In Fig.~\ref{Fig:DelT}, $\Delta(\Gamma,T)$ is given as a function of $T$
for a few $\Gamma$. In addition, $\Delta(\Gamma,T=0)/\Delta_{00}$
is shown in Fig.~\ref{Fig:Tc}, 
which decreases slowly (slower than $T_c/T_{c0}$) with $\Gamma$.

\begin{figure}[ht]
\begin{center}
\epsfig{file=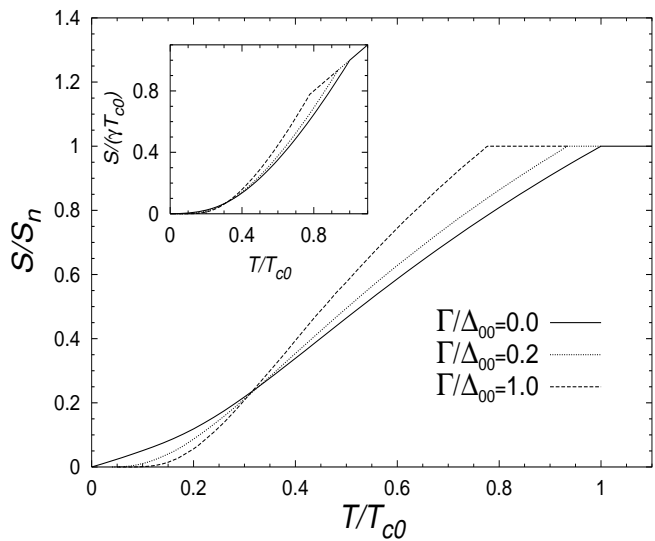,width=8.5cm,height=7cm}
\end{center}
\medskip
\caption{The entropy $S/S_n$ as a function of $T/T_{c0}$ for
a few $\Gamma/\Delta_{00}$. The inset shows 
$S/(\gamma T_{c0})$ vs. $T/T_{c0}$.}
\label{Fig:ent}
\end{figure}

We further calculate the thermodynamics. The entropy $S$
is given by
\begin{eqnarray}
S & = & -2\int_{-\infty}^{\infty} {\rm d}E\ N(E) \left\{f(E)\ln [f(E)]
\right.\nonumber\\
& & \hspace*{3cm} \left. + [1-f(E)]\ln [1-f(E)]\right\}
\end{eqnarray}
with Fermi function $f(E)=1/(1+e^{\beta E})$. In Fig.~\ref{Fig:ent}
we have shown the entropy $S/S_n$ vs. $T/T_{c0}$ for a few $\Gamma$, 
where $S_n=\gamma T$ with $\gamma={2\over 3}\pi^2N_0$ is the entropy in 
the normal state. In the inset the same figure is reploted but $S$ is
scaled by a constant $\gamma T_{c0}$. This is helpful to understand the 
specific heat shown below.

\begin{figure}[ht]
\begin{center}
\epsfig{file=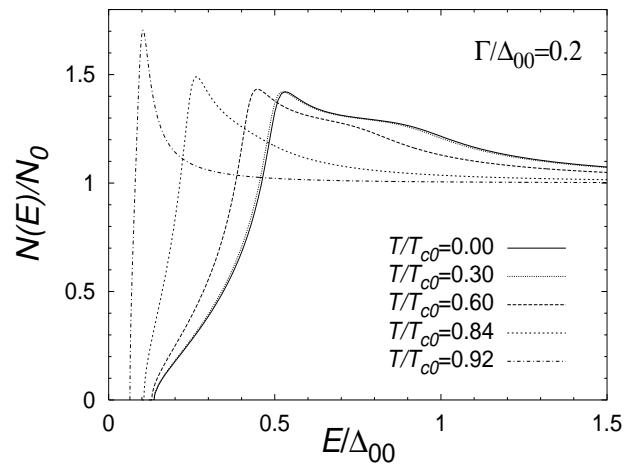,width=8cm,height=6cm}
\end{center}
\medskip
\caption{The DOS with change of $T$ for $\Gamma/\Delta_{00}=0.2$
(with corresponding $T_c/T_{c0}=0.934$). It is seen that
the DOS at $T/T_{c0}=0.3$ is only slightly changed from that
at $T=0$. With increasing $T$, the bump at relatively high energy 
dissolves gradually, and the peak becomes narrow and
simultaneously shifts towards zero. When $T\rightarrow T_c$, the DOS is close to
an inverse square root function valid for an isotropic s-wave gap. Finally it will
become the straight line ``1'' at $T=T_c$.}
\label{Fig:DOST}
\end{figure}

The specific heat is given by $C=T{\partial S\over \partial T}$, which
can be explicitly expressed as
\begin{eqnarray}
{C\over C_n} & = & {3\over 2\pi^2 T^3} \int_{0}^{\infty} {\rm d}E\ 
{N(E)\over N_0} E^2{\rm sech}^2 ({E\over 2T})\nonumber\\
 & & +{6\over \pi^2}\int_{0}^{\infty} {\rm d}E\ 
{\partial \over \partial T} \left[ {N(E)\over N_0} \right]
\left[\ln (1+e^{\beta E})-{\beta E\over 1+e^{-\beta E}}\right]\ ,
\nonumber\\
\label{C}
\end{eqnarray}
where $C_n=\gamma T$ is the specific heat in the normal state.
In Eq.~(\ref{C}) the second term is contributed by the $T$-dependence of 
the DOS. It is important only at high temperatures because at low $T$,
about $T<0.2T_{c0}$ (see Fig.~\ref{Fig:DelT}),
$\Delta(\Gamma,T)$ is nearly $T$-independent and also the DOS. 
For an instructive look at how the DOS changes with $T$ in the whole range,
we have shown in Fig.~\ref{Fig:DOST} an example for $\Gamma/\Delta_{00}=0.2$.

\begin{figure}[ht]
\begin{center}
\epsfig{file=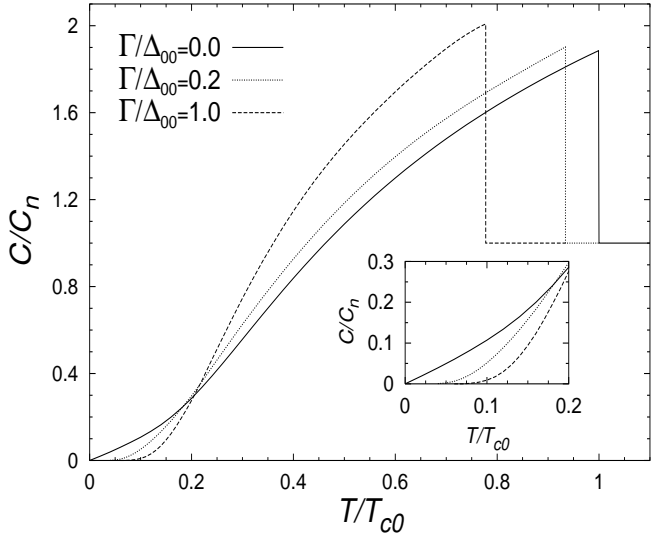,width=8.5cm,height=7cm}
\end{center}
\medskip
\caption{The specific heat $C/C_n$ vs. $T/T_{c0}$ for
$\Gamma/\Delta_{00}=0,\,0.2$ and $1.0$. The low temperature region is
enlarged in the inset.}
\label{Fig:hc}
\end{figure}

The specific heat $C/C_n$ vs. $T/T_{c0}$ is shown in Fig.~\ref{Fig:hc}. 
We mainly concern the low temperature behaviors, which are enlarged 
in the inset. The notable observation is that the $T$-dependence
changes from a power law behavior in the pure case into an activated 
behavior in the presence of impurities. This is due to the opening of 
a full gap by impurities as seen from the DOS. Thus the low energy 
quasi-particle excitations are exponentially small.

The above point becomes more transparent if we present the analytical results 
for the specific heat at low $T$ where only the first integration in 
Eq.~(\ref{C}) contributes. In the pure case, the DOS is linear at low 
energy $E\ll \Delta$: ${N(E)\over N_0}={\pi\over 4}{E\over \Delta}$. 
For about $0.1\Delta/T>5$, i.e., $T<0.05T_{c0}$, this linear formula can be 
used throughout the whole range in the integration. Then we have
\begin{equation}
{C\over C_n} = {27\over 4\pi}\zeta(3){T\over \Delta}\simeq 0.93 {T\over T_{c0}}\ ,
\end{equation}
i.e., a linear $T$-dependence (or $T^2$ dependence for $C$).
When $\Gamma>0$, the lower limit of the integration becomes $\omega_g$. 
In view of ${\rm sech}^2 (E/2T)\simeq 4 \exp(-E/T)$ at low $T$ (about $\omega_g/T>5$), 
an exponential factor $\sim \exp (-\omega_g/T)$ is naturally expected 
after the integration irrespective of the concrete form of $N(E)$.
Explicitly, for $T<\Gamma\ll \Delta_{00}$ we have obtained $(\omega_g \simeq \Gamma)$
\begin{eqnarray}
{C\over C_n} & = & {3 \Gamma^{5/2}\over \sqrt{2}\pi^{3/2}T^{3/2}\Delta}
\left[1-{\Gamma \over \Delta}\ln (1+{\Delta \over \Gamma})\right]^{-1} \nonumber\\
 & & \times (1+{27T\over 8\Gamma}) e^{-\Gamma/T}\ .
\end{eqnarray}

Experimentally the specific heat was measured by Nohara {\it et al.}\cite{Nohara99}
for Y(Ni$_{1-x}$Pt$_x$)$_2$B$_2$C with $x=0$ and $0.2$. By fitting the
experimental data they have exhibited a power law relation $C\propto T^n$ with
$n\sim 3$ for $x=0$ compound and $C\propto \exp (-{\rm const.}/T)$ for
$x=0.2$. The essential features of the experimental data in both pure and 
impure systems have been reproduced by our theoretical results,
although some differences exist for the accurate $T$-dependences 
between theory and experiment.

In addition, the specific heat jump at the transition temperature can be 
analytically derived. For a pure system, we have
\begin{equation}
\left.{C-C_n\over C_n}\right|_{T=T_{c0}}=1.426 
{\langle f^2({\bf k})\rangle ^2\over
\langle f^4({\bf k})\rangle}\simeq 0.885\ .
\end{equation} 
For $\Gamma>0$, the formula is algebraically more complicated:
\begin{eqnarray}
\left.{C-C_n\over C_n}\right|_{T=T_{c}} = 
12 (1+a)^2\left[1-{a\over 1+a}x\psi^{(1)}({1\over 2}+x)\right]^2\times & & 
\nonumber \\
\left\{7\zeta(3)-{b\over 2}\psi^{(2)}({1\over 2}+x)
+3ax^{-1}\left[{\pi^2\over 2}-\psi^{(1)}({1\over 2}+x)\right]\right. + 
& & \nonumber\\
ax\left[x^{-2}\left({\pi^2\over 2}+\psi^{(1)}({1\over 2}+x)-
2x^{-1}[\psi({1\over 2}+x)-\psi({1\over 2})]\right)\right. & & \nonumber\\
\left. \left. \ -{1\over 6}\psi^{(3)}({1\over 2}+x)\right]
\right\}^{-1}\ ,& &\label{Cjump}
\end{eqnarray}
where $x=\Gamma/(2\pi T_c)$, $a=\langle w^2\rangle$, 
$b=\langle w^4\rangle$ with $w=\sin^4\theta\cos 4\phi$, and 
$\psi^{(i)}(z)$ ($i=1,2,3$) are the poly-gamma functions.
From Eq.~(\ref{Cjump}), we obtain
$\left.{C-C_n\over C_n}\right|_{T=T_{c}}=0.901 (1.016)$ for
$\Gamma/\Delta_{00}=0.2 (1)$, which are consistent with the numerical 
values in Fig.~\ref{Fig:hc}.

\section{conclusion}
In the present paper we investigate the impurity effects on s+g-wave
superconductivity in YNi$_2$B$_2$C and possibly LuNi$_2$B$_2$C.
Unlike in the usual nodal superconductors we have discovered that the results
in Born and unitarity limit are practically the same. There is
no resonant scattering associated with nonmagnetic impurities. Also unlike
in the usual nodal superconductors, the quasi-particle energy gap opens up
immediately by impurity scattering and grows quickly with the scattering rate. 
In other words the nodal excitations
are prohibited immediately when impurities are introduced, which is clearly seen
from the quasi-particle density of states.
Indeed we could evaluate that the energy gap opens up even for 
Y(Ni$_{1-x}$Pt$_x$)$_2$B$_2$C with $x=0.01$. From the value for $\Gamma$ at $x=0.05$,
we estimate $\Gamma/\Delta_{00}\sim 0.1$ for the crystal with $x=0.01$,
which is substantial for a gap opening. Therefore the s+g-wave superconductivity is 
very sensitive to impurity scattering.

Also we have calculated the specific heat, which shows a change
from a power law dependence of $T$ in a pure system into an activated 
behavior in the presence of impurities. Compared with the 
experimental data by Nohara {\it et al.},\cite{Nohara99} 
the theoretical results have reproduced the main features. 
The accurate comparison needs improvements in both theoretical and 
experimental side.

Finally we point out that the sensitivity to impurities 
predicted by s+g-model is confirmed by the angular dependent 
thermal conductivity measurements on the single crystal 
Y(Ni$_{1-x}$Pt$_x$)$_2$B$_2$C with $x=0.05$,\cite{Kamata} where the gap
anisotropy has been cleared out.
In a forthcoming paper we shall discuss other properties
such as superfluid density etc. within the same formal treatment.

\section*{acknowledgment}
We thank K. Kamata, K. Izawa and Y. Matsuda for many useful discussions on 
ongoing experiments on YNi$_2$B$_2$C and Pt-substituted samples.
Q. Y. also acknowledges P. Hirschfeld, L. Y. Zhu and J. X. Zhu for helpful 
discussions and J. Goryo for collaboration in the early stage of the work. 
H. W. acknowledges the support by the Korean Science Research Foundation
through Grant No 1999-2-114-005-5.

\end{document}